\documentclass[runningheads,a4paper]{llncs}
\usepackage{subcaption}
\usepackage{xcolor}
\usepackage{hyphenat}
\usepackage[english]{babel}
\defineshorthand{~=}{\hyp{}}

\usepackage{picins}
\usepackage{graphicx}

\usepackage{url}
\usepackage{comment}
\usepackage{amsmath}
\usepackage{booktabs}
\usepackage{multirow}
\usepackage{longtable}
\usepackage{amssymb}
\usepackage{pifont}
\usepackage{cleveref} 
\urldef{\mailsa}\path|{suhrid.gupta1, tawfiqul.islam, rbuyya}@unimelb.edu.au|  
\newcommand{\keywords}[1]{\par\addvspace\baselineskip
\noindent\keywordname\enspace\ignorespaces#1}

\newcommand{\cmark}{\ding{51}}%
\newcommand{\xmark}{\ding{55}}%

\begin{document}

\mainmatter  

\title{Proactive and Reactive Autoscaling Techniques for Edge Computing}
\titlerunning{Proactive and Reactive Autoscaling Techniques for Edge Computing}

\author{Suhrid Gupta \and Muhammed Tawfiqul Islam \and Rajkumar Buyya}
\authorrunning{Gupta, Islam, Buyya}

\institute{Cloud Computing and Distributed Systems (CLOUDS) Lab,\\
School of Computing and Information Systems,\\
The University of Melbourne, Australia\\
\email{\{suhrid.gupta1, tawfiqul.islam, rbuyya\}@unimelb.edu.au}\\
\url{https://clouds.cis.unimelb.edu.au/}}

%
%

\maketitle

\begin{abstract}
Edge computing allows for the decentralization of computing resources. This decentralization is achieved through implementing micro-service architectures, which require low latencies to meet stringent service level agreements (SLA) such as performance, reliability, and availability metrics. While cloud computing offers the large data storage and computation resources necessary to handle peak demands, a hybrid cloud and edge environment is required to ensure SLA compliance. Several auto-scaling algorithms have been proposed to try and achieve these compliance challenges, but they suffer from performance issues and configuration complexity. This chapter provides a brief overview of edge computing architecture, its uses, benefits, and challenges for resource scaling. We then introduce Service Level Agreements, and existing research on devising algorithms used in edge computing environments to meet these agreements, along with their benefits and drawbacks.
\end{abstract}

\def\chaptertitle{Introduction}

\section{\chaptertitle}
\label{ch:introduction}

Cloud computing architectures leverage the on-demand accessibility of the Internet. The applications deployed here utilize the vast resources of the cloud to perform a task and relinquish it once it is complete for the other sub-modules in the application to request~\cite{rimal2009taxonomy}. In the early days, a singular end-point would be used to access these services, however nowadays the architecture is multi-regional allowing effortless access from across the world. This was achieved through the use of content delivery networks (CDN) located in several regions to allow for data to be quickly replicated and served to clients. This architecture model allows for the processing of large-scale data in a near real-time manner.\par

During the early twenty-first century, this architecture paradigm dominated the Information Technology (IT) industry. Compared to traditional monolithic architectures, the ease of deployment, and scalability, coupled with the economic benefits ensured its dominance. The increasing popularity of hand-held devices as well as home appliances has resulted in data being largely produced at the edge of the cloud network. Thus, processing this large amount of data solely on the cloud proved to be an inefficient solution due to the bandwidth limitations of the network~\cite{shi2016edge}. To counteract this inefficiency, edge computing paradigms were built on the previous foundation of CDNs~\cite{satyanarayanan2017emergence}. Edge computing architectures ensure data processing services and resources exist at the peripheries of the network~\cite{cao2020overview}. The architecture extends and adapts the computing and networking capabilities of the cloud to meet real-time, low latency, and high bandwidth requirements of modern agile businesses.\par

\begin{figure}[htb]
    \centering
    \caption{Overview of edge computing architecture}
    \includegraphics[width=0.9\linewidth]{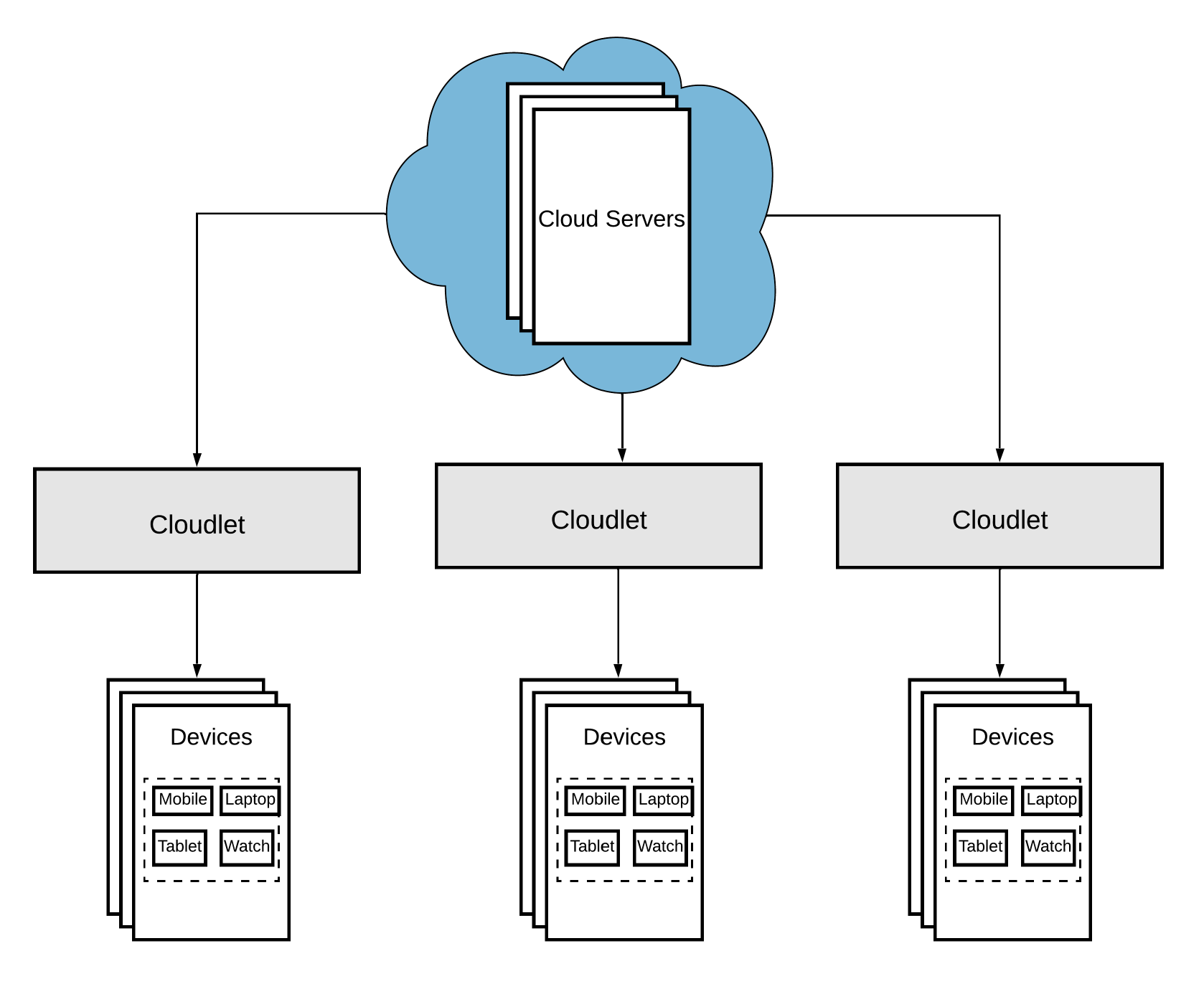}
    \label{fig:edge-architecture-overview}
\end{figure}

\textcolor{green}{Edge computing} deploys several lightweight computing devices known as \textit{cloudlets} to form a ``mini-cloud'' and places them in close proximity to the end-user data~\cite{liu2019survey}. This reduces the latency in terms of client-server communication and data processing. Figure~\ref{fig:edge-architecture-overview} shows a high-level overview of this architecture. Cloudlets can also be easily scaled depending on the resource requirements per edge architecture~\cite{ren2019survey}. However, due to the dynamic resource requirements which may fluctuate from time to time, the resources allocated to cloudlets must be dynamically scaled too. This dynamic scaling, along with the inherent latency present between the cloud layer and the edge cloudlets, poses a significant problem to real-time resource scaling~\cite{varghese2016challenges}.\par

One method of mitigating this scaling latency is through the use of \textcolor{green}{micro-service} applications. By employing a micro-service architecture, the resources in a cloudlet are distributed as a collection of smaller deployments that are both independent and loosely coupled~\cite{villamizar2015evaluating}. This loose coupling ensures that parts of the cloudlet can be scaled as required, further reducing the time required to scale resources as compared to scaling the cloudlet monolithically.\par

The scaling of these micro-service resources is done automatically through a process known as auto-scaling. While most container orchestration platforms come bundled with default auto-scaling solutions, and these solutions are sufficient for most applications, they fall apart when scaling resources for time-sensitive services processing real-time data such as the ones used in healthcare require stringent compliance to service level agreements (SLA) on metrics such as application latency. This has led to further research on auto-scaling solutions for edge computing applications. These primarily fall into two categories. Reactive auto-scaling solutions attempt to modify the micro-service resource allocation once the required resources exceed the current allocation. These algorithms are simple to develop and deploy, however, the time taken to scale resources leads to a degradation of resource availability and violates \textcolor{green}{SLA compliance}~\cite{podolskiy2018iaas}. To counteract these pitfalls, proactive auto-scaling solutions attempt to model resource allocation over time and effectively predict the resource requirements. By doing so, the micro-service resources can be scaled in advance through a process known as ``cold starting''. This approach removes the latency inherent in scaling resources, however, the algorithms are extremely complex to develop, train, and tune to specific edge applications~\cite{straesser2022not}.\par

\def\chaptertitle{Background and Key Concepts}

\section{\chaptertitle}
\label{ch:background}

In this section, a brief introduction to Service Level Agreements is provided, followed by an overview of micro-service architectures. Finally, we give a brief description of the architecture of \textcolor{green}{Kubernetes} and its \textcolor{green}{auto-scaling} algorithms.

\subsection{Service Level Agreements}
\label{sec:ch2-sla}

\begin{figure}[htb]
    \centering
    \caption{Overview of service level agreements}
    \includegraphics[width=0.9\linewidth]{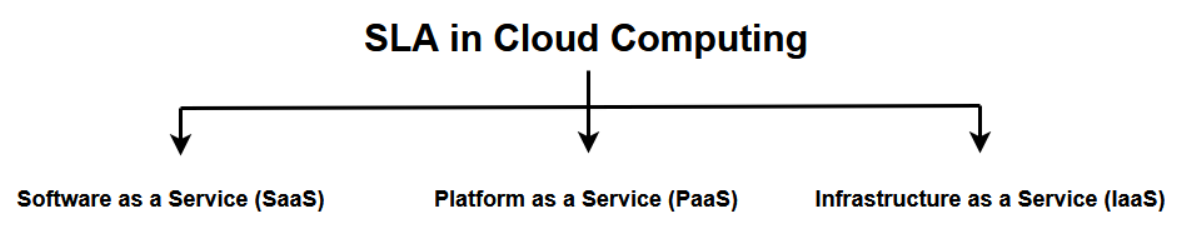}
    \label{fig:sla-types}
\end{figure}

Cloud computing generally exposes resource using a pay-as-you-go service. These lucrative plans have led to the implementation of applications and hardwares being delivered as Software as a Service (SaaS), Platform as a Service (PaaS), and Infrastructure as a Service (IaaS), as shown in Figure \ref{fig:sla-types}. However, consumers of such services have demands which may vary significantly, and it is impossible to fulfill all these expectations. Thus a balance needed to be struck in order to commit to an agreement~\cite{patel2009service}. \par
This commitment is known as a Service Level Agreement (SLA). This SLA defines the expected services provided by the provider, and agreed to by the consumer. For example, one of the most common metric by which SLAs are negotiated between providers and consumers is the availability of service.

\subsubsection{Availability of Services}
\label{subsec:ch2-svc-availability}
Availability is defined to ensure that the functional performance of the edge deployment is maintained for an agreed period. SLAs mostly define either monthly or yearly downtime in order to compute service credits for billing purposes~\cite{mirobi2015service}. The downtime can be calculated using the formulae:
\begin{equation}
    downtime_{monthly} = \cfrac{100 - Availability\%}{100} \times 30 \times 24
\end{equation}

\begin{equation}
    downtime_{yearly} = \cfrac{100 - Availability\%}{100} \times 365
\end{equation}

Table~\ref{table:sla-availability} shows the expected down-times for several SLA availability percentages.

\begin{table}
    \caption{Summary of SLA availability}\label{table:sla-availability}
    \centering
    \begin{tabular}{rcl}
        \toprule
        \textbf{Availability \%} & \textbf{Monthly Downtime} & \textbf{Yearly Downtime}\\
        \midrule
        90\% & 72 hours & 36.5 days\\
        99\% & 7.2 hours & 3.65 days\\
        99.9\% & 43.8 minutes & 8.76 hours\\
        99.99\% & 4.38 minutes & 52.56 minutes\\
        99.999\% & 25.9 seconds & 5.26 minutes\\
        \toprule
    \end{tabular}
\end{table}

\subsection{Microservice Architecture}
\label{sec:ch2-micro-svc-arc}

\begin{sloppypar}
Micro-service architectures involve decomposing an application into several loosely coupled services, and deploying them on separate cloudlet servers known as ``nodes''. These services communicate with each other through a lightweight framework such as RESTful APIs~\cite{li2021understanding}. Within these services, application data and commands are stored and executed within ``containers''. Typically, these architectures provide scalability, as well as ease of deployment and modification. Availability however, remains an important concern for such deployments. For a deployment to be classified as ``highly available'', it must be accessible at least 99.999\% of the time. For example, a highly available search engine would only face 5 minutes of down time per year~\cite{nabi2016availability}. Therefore, an orchestration mechanism is required to manage the deployment and communication of these containers.\par
\end{sloppypar}

\begin{figure}
    \centering
    \caption{Features of container orchestration}
    \includegraphics[width=0.9\linewidth]{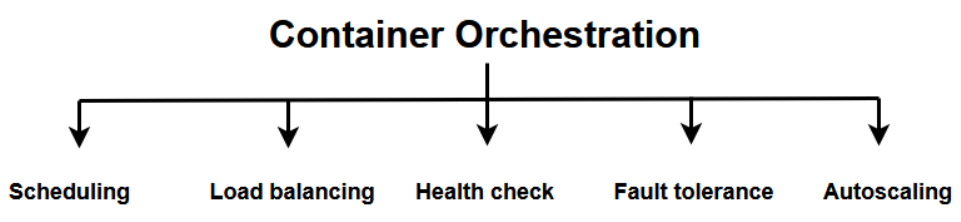}
    \label{fig:container-orchestration}
\end{figure}

\textcolor{green}{Container orchestration} allows the micro-service application to customize how the deployment, monitoring, and controlling functions~\cite{casalicchio2019container}. Figure~\ref{fig:container-orchestration} depicts the typical features of container orchestration.\par
\textit{Scheduling} defines the rules on the number of containers to be executed at any given time. Scheduling also places containers on specific nodes based on availability and best performance.\par
\textit{Load balancing} distributes the resource usage among multiple micro-service nodes. By default, a round-robin policy is implemented, although more complex policies may be implemented at the discretion of the developer.\par
\textit{Health checks} ensure that the container is still capable of responding to queries. Typically, these are done using a periodic light-weight HTTP request and verifying the response.\par
\textit{Fault tolerance} maintains several replicas of containers, a strategy commonly used to achieve the high availability mentioned above. Health checks are used to ensure the replicas are functioning, and they typically have strategies to ensure there is no mismatch in data between two fault tolerant containers.\par
\textit{Autoscaling} is the process of automatically adding or removing resources or containers. Internal metrics such as CPU usage are typically used, however custom policies can also be implemented at the discretion of the developer.\par

\subsection{Kubernetes Architecture}
\label{subsec:ch2-k8s-overview}

\begin{figure}
    \centering
    \caption{Overview of Kubernetes architecture}
    \includegraphics[width=0.5\linewidth]{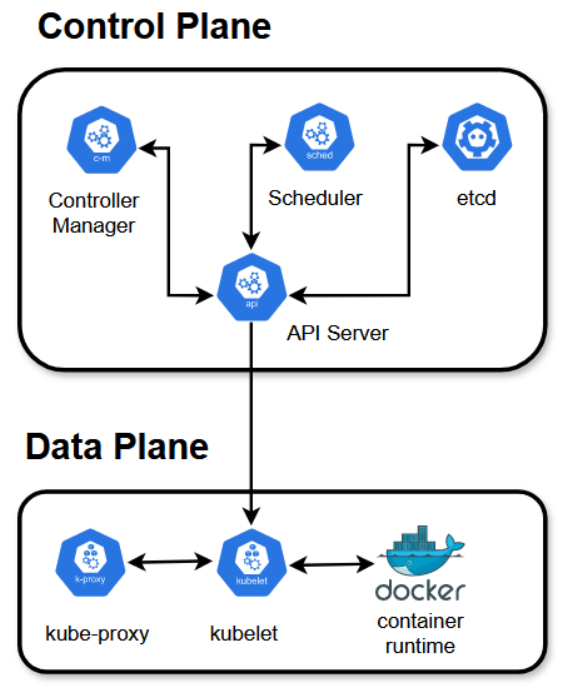}
    \label{fig:k8s-arch}
\end{figure}

Kubernetes is one of the most popular open-source container orchestration platforms~\cite{vayghan2021kubernetes}. Initially referred to as ``Borg'', the project was used internally at Google to deploy the majority of their cloud applications before becoming an open-source application~\cite{burns2016borg}. Figure~\ref{fig:k8s-arch} shows the high-level architecture. The Kubernetes deployment has a controller / worker architecture. The nodes in the Kubernetes cluster are split into either \textit{control plane nodes} and \textit{data plane nodes}. The \textit{control plane nodes} have a collection of processes which help monitor and maintain the desired state of the deployment. The \textit{data plane nodes} contain processes which run the containers doing the actual work, and are managed by the control plane.\par
The smallest unit of work in a Kubernetes deployment is known as a \textit{pod}~\cite{baier2017getting}. This is a collection of containers sharing an IP address and port. In summary, micro-service architectures are said to be containerized and deployed on Kubernetes in the form of pods~\cite{vayghan2021kubernetes}.\par

\subsubsection{Control Plane}
\label{subsubsec:ch2-k8s-control-plane}
The \textit{API Server} is the primary communication endpoint for the entire deployment. Every component in the architecture communicates through it to exchange information. It is also used to update the current deployment state. The API Server is a simple RESTful API implementation, exposing well-documented APIs for access by other components as well as developers. Multiple replicas of this component are typically maintained to ensure high availability.\par

The \textit{etcd} is a data store which persists the deployment state in a key-value format. The data is serialized unlike in the stateless API server. This data adheres the properties of \textit{recovery} and \textit{availability}. \textit{Recovery} ensures that any corruption of data is reverted using a system of backups such as checkpoints. \textit{Availability} ensures that the deployment is reachable by the end-user regardless of the traffic being requested on the network.\par

\textit{Controller Manager} implements the desired deployment state. During initial deployment, the controller manager inputs the required workload as the desired state, after which it continually monitors the deployment state using a system of looping controls. If the deployment requires modifications, they are achieved using the API server, and the deployment is brought back into alignment with the desired state.\par

Finally, the \textit{scheduler} decides the location where the pod will be deployed. The scheduler runs a control loop which searches for unscheduled pods using the API server. It then assigns the pods to a dataplane node based on several predicates and priorities such as resource requirements and node affinity respectively.

\subsubsection{Data Plane}
\label{subsubsec:ch2-k8s-data-plane}

The \textit{container runtime} is a process which downloads or ``pulls'' the image for the required container onto the node. Kubernetes supports a wide range of runtimes, but some of the popular solutions are CRI-O, containerd, and Docker.\par

The most important process running on every data plane node is the \textit{kubelet}. This process executes the image assigned to the node via the container runtime, perform health checks, and reports the node status to the control plane.\par

Another data plane process is the \textit{kube-proxy}, which manages the rules for forwarding requests to services, as well as the IP tables of nodes. If a service is added or removed, kube-poxy updates the IP table accordingly.\par

\subsection{Auto-scaling Overview}
\label{sec:ch2-auto-scaling}

The process of scaling nodes, pods, or other resources depending on requirements in an automated manner is known as \textit{auto-scaling}. Typically, there are three variations of auto-scaling.\par

\textit{Cluster auto-scaling} modifies the number of nodes running in the entire deployment, or cluster. Dynamically allocating nodes based on resource requirements helps to manage the cost of running the deployment. The autoscaler works by looping through two tasks. The first watches for unscheduled pods, the second checks if the current deployed pods can be merged on a smaller number of nodes.\par

\textit{Vertical pod auto-scaling} modifies the CPU and memory resources assigned to pods. By default, the scheduler reserves a larger amount of these resources to pods than is usually required. By performing vertical pod auto-scaling, the cluster can better manage its over-provisioned resources in real-time.\par

\textit{Horizontal pod auto-scaling} is the most commonly used auto-scaling strategy~\cite{baresi2021kosmos}. It modifies the number of pods assigned to a task, based on the resources being requested. A periodic control loop compares the actual resource utilization with the target utilization defined by the initial deployment setup, and scales the number of pods accordingly.

\section{Edge Computing Implementation}
\label{sec:ch3-edge-implementation}
\begin{sloppypar}
Yu \textit{et al}.~\cite{yu2017survey} states when implementing edge computing architectures, researchers typically focus on two models, namely the hierarchical model, and the software-defined model.\par
\end{sloppypar}
\subsection{Hierarchical Model}
\label{subsec:ch3-hierarchical-model}

Since edge and cloud servers are deployed at varied distances from the end users, it makes sense that the edge architecture is to be divided into such a hierarchy, with each layer defining the functions relative to resource availability and distance.\par

Smeliansky~\cite{smeliansky2018hierarchical} gives an overview of such a hierarchical model. The bottom tier is usually comprised of a geo-distributed servers which receive workload from mobile devices via wireless links. These servers are then connected to a higher tier level of servers which comprise of the remote data centers. If the mobile workload received by a lower tier server exceeds its computational capacity, it can offload this excess workload to a higher tier server. In this way, the architecture can serve large load peaks.\par

Several research efforts exist for such a model. Tong \textit{et al}.~\cite{tong2016hierarchical} proposed an edge cloud hierarchy which could be used to serve high load demands from mobile users. In such a model, the cloudlet servers were deployed at the edge layer, while the cloud established as a tree hierarchy. By using such an implementation, the edge layer was able to aggregate its servers resource abilities to better serve peak workloads.\par

Jarwah \textit{et al}.~\cite{jararweh2016future} demonstrated a similar approach to a hierarchical model, which integrated Mobile Edge Computers (MEC) servers and cloudlets. The mobile users obtained specific services as per requested, and the MEC served the ability to deliver their computational and storage requirements.\par

\subsection{Software-defined Model}
\label{subsec:ch3-software-defined-model}

The number of end users and devices which connect to the edge architecture typically numbers in the millions~\cite{yu2017survey}. Thus, management of an application deployed on this architecture can prove to be significantly challenging. To address these complexities, Software Defined Networks (SDN) were proposed.\par

According to Wang \textit{et al}.~\cite{wang2017controller}, SDNs distinguish themselves from conventional networking approaches by decoupling the control plane from the data plane. The control plane is constructed using a combination of dedicated controllers. These serve as the control center of the SDN model, while the data-plane simply forwards data packets in the form of a network switch. Such a decoupling makes the architecture highly flexible, as well as simplifies network management~\cite{liu2015device}. However they have drawbacks which include performance issues if not developed properly.\par

Despite these challenges, there have been a number of research efforts based on the SDN model. Du and Nakao~\cite{du2016application} presented an MEC model which was application specific. In this model, the software-defined data plane acts as a Mobile Virtual Network Operator (MVNO). The mechanism computes a hop-count based tethering as well as optimizes the process.  Fairness among determining user resources is determined through regulating concurrent TCP connections.\par

Similarly, Jaraweh \textit{et al}.~\cite{jararweh2016sdmec} proposed an SDN model which integrated its capabilities to the MEC system. The management and administrative requirements for the entire model could be reduced in this manner.\par

Finally, Salman \textit{et al}.~\cite{salman2015edge} demonstrated an integration of SDN, MEC, and a Network Function Virtualization (NFV)m which was capable of achieving better MEC performance in mobile networks than the other two works. Such a model could be further extended to enable an IOT-capable deployment.\par

\section{Edge Computing Issues and Challenges}
\label{sec:ch3-edge-issues}

Maintaining SLA compliance in an edge computing architecture poses several unique challenges. In this section, we briefly describe a few main issues.

\subsection{Resource Allocation}
\label{subsec:ch3-edge-resource-alloc}

Cao \textit{et al}.~\cite{cao2020overview} demonstrated the key differentiations traditional cloud computing architectures have compared to edge architectures, while asserting that edge deployments remain an extension of the cloud. The aim of cloud computing infrastructures is to process huge amounts of data from multi-regional zones, or in the best case, globally. This is done so as to perform in-depth analysis in diverse fields such as health-care, robotics, and business decision making. Traditionally, they also dealt with non-real-time data for decision-making~\cite{premsankar2018edge}. On the other hand, edge computing usually handles smaller scale data, locally clustered and isolated in separate zones, and highly real-time in nature~\cite{mishra2020early}. The data processed in traditional cloud computing environments are also generally done using a high network bandwidth. This is due to the large distances data needs to be transmitted over to reach the data centres and cloud servers. Such data transmission places an enormous burden on the cloud network, and poses multiple security challenges in ensuring that the data is not compromised in transit.\par

The real-time nature of edge computing applications necessitates a method of resource allocation which ensures minimal cost of deployment, and maximum efficiency in terms of performance. Micro-service container orchestration technologies are leveraged to achieve these aims. Kristiani \textit{et al}.~\cite{kristiani2019} demonstrated an edge computing architecture, where the edge layer consists of Kubernetes nodes. Such a deployment increases the scalability, as well as maintains the ease of deployment, upgrade, and removal of nodes in the \textcolor{green}{edge layer}. Scaling of resources through the means of auto-scaling depending on the resource requirements is crucial to the architecture's performance. Default solutions such as the inbuilt autoscaler provided by Kubernetes, while generally useful for cloud applications, are unsuitable for edge architectures according to Phan \textit{et al}.~\cite{phan2022traffic}. They note that due to the algorithm's default nature to allocate resources in a round-robin manner, they do not take into account which Kubernetes nodes require priority resources allocation, violating edge architecture paradigms.

\subsection{Cold Start Problem}
\label{subsec:ch3-cold-start}

To explain the \textcolor{green}{cold start} problem, we use the example of horizontal pod auto-scaling in Kubernetes. When the control plane requests for a deployment replica to be scaled up, Kubernetes adds more pods to the data plane nodes.  Based on the internal workload, the pod needs to be elastically scaled out~\cite{beni2021reducing}. Even though the pod start up time is significantly quicker than, say, a traditional virtual machine, there is a latency inherent to bootstrapping the container, preparing the pod environment based on the deployment specification, and initialising the code present in the container image, and registering the pod in a ``ready'' state to the Kubernetes control plane.\par

Several techniques exist to mitigate this \textcolor{green}{resource latency}. The Kubernetes container runtime uses snapshots~\cite{cadden2019seuss}, lazy fetching of container images~\cite{lorenzo2019fogdocker}, and container queues~\cite{lin2019mitigating}. However, these measures do not eliminate the issue of the inherent latency in installing and registering the resource. Due to this, researchers looked into scaling resources in a predictive manner, so as to ensure the micro-service application has enough time to spool up resources before the actual demand comes in. This process, by which resources are created and registered with the container orchestration before the expected workload is known as resource \textit{cold start}.\par

\subsection{SLA Guarantees}
\label{subsec:ch3-sla-edge}

There are several challenges posed in providing SLA guarantees in an edge deployment:
\begin{itemize}
    \item Users queueing for large periods of time to use a service~\cite{venticinque2011cloud}.
    \item Degradation of application performance due to peak levels of workload, leading to user dissatisfaction~\cite{sakr2012sla}.
    \item Incorrect resources being allocated to the application, leading to either a degradation of availability, or large cost of application deployment~\cite{houlihan2014auditing}.
\end{itemize}

Several strategies have been proposed to counteract these challenges. Linlin \textit{et al}.~\cite{wu2013sla} proposed a customer-driven strategy  to minimize the provisioning costs. The algorithm considers the customer profiles as well as cloud providers' quality parameters such as response time to dynamically handle customer requests. Rajkumar \textit{et al}.~\cite{rajavel2012achieving} proposed a solution for alleviating the issue of delay in service allocation to users through the use of a novel hierarchical scheduling algorithm. This algorithm increases the performance of the scheduling algorithm, thus reducing the wastage of resources, and minimizing wait times. Sakr \textit{et al}.~\cite{sakr2012sla} introduced a novel approach to combat application performance degradation by using a middleware between consumers and the cloud. This middleware helps to facilitate dynamic provisioning of cloud databases based on consumer requirements, tailoring their needs and requirements to mitigate peak usages being hit often.

\section{Reactive Autoscaling Strategies}
\label{sec:ch3-reactive-solutions}

\begin{table}
    \caption{Summary of reactive auto-scaling solutions}\label{tab:reactive-autoscalers}
    \centering
    \begin{tabular}{ ccccccccc }
         \toprule
         \multirow{2}{*}{\textbf{Features}}&\multicolumn{8}{c}{\textbf{Reactive algorithms}}\\
         \cmidrule{2-9}
         &\cite{phan2022traffic}&\cite{kampars2017auto}&\cite{zhang2019quantifying}&\cite{srirama2020application}&\cite{hoenisch2015four}&\cite{santos2020qoe}&\cite{sheganaku2023cost}&\cite{taherizadeh2019dynamic}\\
         \midrule
         \multicolumn{1}{r}{Simple parameter tuning} & \cmark & \xmark & \cmark & \xmark & \cmark & \cmark & \cmark & \xmark\\
         \multicolumn{1}{r}{Custom metrics} & \xmark & \xmark & \xmark & \xmark & \xmark & \cmark & \cmark & \xmark\\
         \multicolumn{1}{r}{Light-weight deployment} & \cmark & \cmark & \cmark & \cmark & \xmark & \xmark & \xmark & \cmark\\
         \multicolumn{1}{r}{Edge architecture compliant} & \cmark & \xmark & \cmark & \cmark & \cmark & \cmark & \cmark & \cmark\\
         \multicolumn{1}{r}{SLA-compliant} & \xmark & \xmark & \xmark & \xmark & \xmark & \xmark & \xmark & \xmark\\
         \toprule
    \end{tabular}
\end{table}

Nunes \textit{et al}.~\cite{nunes2021state} stated that horizontal pod auto-scaling using a reactive strategy remains the most popular auto-scaling technique, as well as research topic. These strategies, despite having limitations such as a reliance on predetermined resource thresholds and a delay in resource scaling, have been popular in research articles.  Dogani \textit{et al}.~\cite{dogani2023auto} stated that this was due to the simplicity and user-friendliness in developing them.\par

Kampars and Pinka~\cite{kampars2017auto} proposed a \textcolor{green}{reactive auto-scaling} algorithm for edge architectures based on open-source technologies. The algorithm scales in a non-standard approach, considering real-time adjustments in the application logic to determine the strategy of scaling, resulting in several improvements in performance.\par

Zhang \textit{et al}.~\cite{zhang2019quantifying} presented an algorithm for determining edge elasticity through container-based auto-scaling. The authors posit that elasticity is a key factor of how an edge deployment as well as the lightweight containers which make up the edge layer perform. The framework not only autoscales container resources, but also monitors resource usage. They were able to show experimentally that to balance system stability with a decent elasticity required careful tuning of parameters such as the cooldown periods of scaling.\par

Srirama \textit{et al}.~\cite{srirama2020application} investigated an container-aware auto-scaling solution which deploys applications to containers which it deems ``best-fit''. The algorithm also uses a rule-based policy to minimize the deployment time, thus mitigating the issue of cold-start. Finally, a dynamic bin-packing sub-algorithm ensures that the applications are deployed on the least required physical servers, thus minimizing wastage of computing resources. The authors experimentally demonstrated that this algorithm minimized  the processing time, cost, and resource utilization.\par

Hoenisch \textit{et al}.~\cite{hoenisch2015four} implemented a four-fold auto-scaling strategy for containerised applications which asks if the containers or servers can be autoscaled horizontally or vertically. This question is formalized as a multi-objective optimization problem, and the approach used reduced the cost of each request by more than 20\%.\par

Santos \textit{et al}.~\cite{santos2020qoe} implemented a quality of experience based auto-scaling of containerized edge deployments. The algorithm can autoscale both horizontally and vertically on a set of quality metrics which can be customized by the end-user. The authors explained that the experimental results displayed a performance comparable to other reactive solutions.\par

Sheganaku \textit{et al}.~\cite{sheganaku2023cost} devised an container-based auto-scaling solution which allocates resources in a four-fold manner similar to Hoenisch \textit{et al}.~\cite{hoenisch2015four}. The authors formulated the problem as a multi-objective optimization problem and applied a Mixed-Integer Linear Programming (MILP) approach to allocate resources to containers. Such an approach demonstratively reduced costs while maintaining SLA constraints.\par

Taherizadeh and Stankovski~\cite{taherizadeh2019dynamic} proposed a multi-level auto-scaling solution using a rule-based approach. The algorithm uses dynamically changing thresholds based on both the container infrastructure as well as application, resulting in improved performance as compared to other reactive approaches.\par

Phan \textit{et al}.~\cite{phan2022traffic} proposed a reactive auto-scaling solution for edge deployments for IoT devices which dynamically allocates resources based on incoming traffic. This traffic-aware horizontal pod autoscaler (THPA) operates on top of the underlying Kubernetes architecture. As discussed above, the default Kubernetes horizontal pod autoscaler scales resources in a round-robin manner, not taking into context which nodes are receiving the highest resource requests. THPA alleviates this issue by modelling the resource requests per Kubernetes nodes. It then intelligently allocates pods to the nodes with higher number of requests. The authors were able to experimentally demonstrate that following such an approach provided a 150\% improvement in response time and throughput. Table \ref{tab:reactive-autoscalers} provides a comparative summary of reactive autoscaling solutions.

\section{Proactive Autoscaling Strategies}
\label{sec:ch3-proactive-solutions}

\begin{table}
    \caption{Summary of proactive auto-scaling solutions}\label{tab:proactive-autoscalers}
    \centering
    \begin{tabular}{ ccccccc }
         \toprule
         \multirow{2}{*}{\textbf{Features}}&\multicolumn{6}{c}{\textbf{Proactive algorithms}}\\
         \cmidrule{2-7}
         &\cite{ju2021proactive}&\cite{meng2016crupa}&\cite{imdoukh2020machine}&\cite{messias2016combining}&\cite{abdullah2020burst}&\cite{alidoost2023introducing}\\
         \midrule
         \multicolumn{1}{r}{Simple parameter tuning} &      \xmark & \xmark & \xmark & \cmark & \xmark & \xmark\\
         \multicolumn{1}{r}{Custom metrics} &               \cmark & \cmark & \xmark & \xmark & \xmark & \xmark\\
         \multicolumn{1}{r}{Light-weight deployment} &      \xmark & \xmark & \xmark & \cmark & \xmark & \xmark\\
         \multicolumn{1}{r}{Edge architecture compliant} &  \cmark & \xmark & \xmark & \xmark & \xmark & \xmark\\
         \multicolumn{1}{r}{SLA-compliant} &                \xmark & \cmark & \cmark & \cmark & \xmark & \xmark\\
         \toprule
    \end{tabular}
\end{table}

Lorido \textit{et al}.~\cite{lorido2014review} showed that compared to reactive algorithms, \textcolor{green}{proactive algorithms} achieved better resource allocation once they had been carefully optimized. Machine learning (ML) techniques such as auto-regressive integrated moving averages (ARIMA) and long short-term memory (LSTM) have gained populary in time-series analysis due to their relative ease of building and efficiency compared to other ML models. Through the careful use of these models, linear patterns in the data can be automatically identified in a short amount of time with relatively constrained resources. There are however several challenges when implementing a proactive algorithm. Time-series analysis models may struggle when dealing with highly complex and non-linear data~\cite{dogani2023auto}. The development of a generalized algorithm for several edge architectures remains a costly process. One of the biggest challenges is the initial lack of training data. Another issue is the exploding or vanishing gradient problem~\cite{pascanu2013difficulty}, though modern algorithms ensure that they avoid this pitfall~\cite{hochreiter2001gradient}. Despite these challenges, their application in scaling of resources with semi-predictable data series remain valuable.\par

Ju \textit{et al}.~\cite{ju2021proactive} presented a proactive horizontal pod auto-scaling solution for edge computing paradigms. The algorithm, known as Proactive Pod Autoscaler (PPA) was designed to predict resource requests on multiple user-defined metrics, such as CPU request and I/O traffic requests. The algorithm does not use any specific machine learning model for the time-series analysis, instead the model is to be inputted by the user. This model agnostic architecture allows for a very high level of customization. The user can deploy an ARIMA, LSTM, or even Bayesian confidence models. In a confidence model, the autoscaler will only deploy resources if the confidence value is seen to be above a specified user-defined threshold. The authors validated their findings by testing the architecture using LSTM and ARIMA models, the results concluded that this algorithm significantly outperformed both the default Kubernetes autoscaler, as well as existing reactive auto-scaling solutions.\par

Meng \textit{et al}.~\cite{meng2016crupa} created a proactive auto-scaling algorithm for forecasting the Kubernetes CPU usage of containers using a time-series prediction. They achieved this using the ARIMA model. The time-series was split into a training and validation set using a 5:1 ratio before being passed to the deep learning model. The authors were able to demonstrate experimentally that such an architecture reduced the forecast errors to 6.5\%, as compared to the baseline of 17\%.\par

Imdoukh \textit{et al}.~\cite{imdoukh2020machine} proposed a proactive auto-scaling solution using an LSTM model, designed for edge computing architectures. The algorithm uses an LSTM neural network to predict future network traffic workload to determine the resources to assign to edge nodes ahead of time (cold-start). The authors experimentally demonstrated that their algorithm was as accurate as existing ARIMA-based proactive solutions, but theirs significantly reduced the prediction time, as well as computed the minimum resource allocation required to handle future workload.\par

Messias \textit{et al}.~\cite{messias2016combining} created a proactive autoscaler using genetic algorithms (GA). The genetic algorithm combines several time-series forecasting models, while having the benefit of not requiring a training phase as the model adapts to the incoming data. The experimental results concluded that this approach produces results comparable to several state-of-the-art proactive models, and can adapt to various time series models.\par

Abdulla \textit{et al}.~\cite{abdullah2020burst} devised an auto-scaling solution which is capable of detecting sudden bursts in dynamic workloads. The algorithm achieves this through a method of workload and resource prediction to make a scaling decision. Experimenting on several burst-heavy workloads, the autoscaler demonstrated significant improvements compared to other state-of-the-art methods.\par

Alidoost \textit{et al}.~\cite{alidoost2023introducing} proposed a workload classification model using a Support Vector Machine (SVM). The algorithm extracts the user's workload characteristics, and then trains the SVM on it. The authors demonstrated a 10\% forecast error reduction compared to other machine learning proactive forecast approaches. 
Table \ref{tab:proactive-autoscalers} provides a comparative summary of proactive autoscaling solutions.\par

\section{Hybrid Autoscaling Strategies}
\label{sec:ch3-hybrid-solutions}

\begin{table}
    \caption{Summary of hybrid auto-scaling solutions}\label{tab:hybrid-autoscalers}
    \centering
    \begin{tabular}{ cccccc }
         \toprule
         \multirow{2}{*}{\textbf{Features}}&\multicolumn{5}{c}{\textbf{Hybrid algorithms}}\\
         \cmidrule{2-6}
         &\cite{xu2007use}&\cite{lama2009efficient}&\cite{ramperez2021flas}&\cite{biswas2017hybrid}&\cite{singh2021rhas}\\
         \midrule
         \multicolumn{1}{r}{Simple parameter tuning} &      \cmark & \cmark & \cmark & \xmark & \xmark\\
         \multicolumn{1}{r}{Custom metrics} &               \cmark & \xmark & \xmark & \cmark & \cmark\\
         \multicolumn{1}{r}{Light-weight deployment} &      \cmark & \xmark & \cmark & \xmark & \xmark\\
         \multicolumn{1}{r}{Edge architecture compliant} &  \xmark & \xmark & \xmark & \xmark & \xmark\\
         \multicolumn{1}{r}{SLA-compliant} &                \xmark & \xmark & \cmark & \cmark & \cmark\\
         \toprule
    \end{tabular}
\end{table}

All the approaches mentioned in \cref{sec:ch3-reactive-solutions,sec:ch3-proactive-solutions} have their benefits and drawbacks. Thus, \textcolor{green}{hybrid} solutions which merge multiple auto-scaling methods were proposed~\cite{qu2018auto}. While hybrid algorithms for cloud-based deployments exist, integrating them into edge architectures pose several challenges due to the lower data storage and computational capacity of the edge layer. Furthermore, extracting the proactive time-series analysis to the cloud layer poses further challenges due to the inherent latency present between the two layers.\par

In 2007, one of the first hybrid algorithms for a distributed deployment was proposed by Jing \textit{et al}.~\cite{xu2007use}. This algorithm combined rule-based fuzzy inference with machine learning forecasting for dynamic resource allocation. The authors experimentally verified their algorithm through a prototype to demonstrate that it can reduce the resource consumption on resource management systems compared to their default resource allocation algorithms.\par

Based on this work, Lama and Zhou~\cite{lama2009efficient} proposed a resource provisioning algorithm for multi-cluster set ups using a hybrid autoscaler. The autoscaler comprised of a combination of fixed fuzzy rule-based logic and a self adaptive algorithm which dynamically tuned the scaling factor. The authors tested this algorithm on a simulation to demonstrate performance benefits compared to existing approaches.\par
\begin{sloppypar}
A hybrid approach for cloud computing architectures was proposed by Ramp{\'e}rez \textit{et al}.~\cite{ramperez2021flas}. The algorithm which was called Forecasted Load Auto-scaling (FLAS), combines a predictive model for forecasting time-series resources, while the reactive model estimates other high-level metrics and delegates for the proactive model, reducing the potential forecast error when encountering previously unseen workloads. The approach was shown to demonstrate efficient resource allocation as compared to other state-of-the-art solutions.\par
\end{sloppypar}
Biswas \textit{et al}.~\cite{biswas2017hybrid} presented a hybrid algorithm designed for cloud computing deployments with service level agreements. The proactive algorithm involves a machine-learning based approach using an SVM model, alongside the reactive algorithm to dynamically allocate resources. The algorithm was experimentally shown to perform better than a pure reactive or proactive solution in most cases.\par

Finally, another cloud computing based autoscaler with SLA-constraints was proposed by Singh \textit{et al}.~\cite{singh2021rhas}. The robust hybrid autoscaler (RHAS) was designed particularly for web applications. The reactive rule-based autoscaler deals with current workload requirements, while the proactive model attempts to predict incoming resource workloads. The proactive forecaster uses a modification of the ARIMA machine learning model, known as the Technocrat ARIMA and SVR Model (TASM)~\cite{singh2019tasm}. The authors tested their algorithm on two data-sets belonging to ClarkNet and NASA. The technique was demonstrated to both reduce cloud deployment cost and SLA violations while giving consistent CPU utilization.
Table \ref{tab:hybrid-autoscalers} provides a comparative summary of hybrid autoscaling solutions.\par

\section{Experimental Case Study}
We propose a novel hybrid auto-scaling algorithm for SLA-constrained edge computing applications, combining a machine learning-based proactive forecaster with a reactive autoscaler for real-time adjustments. Integrated into Kubernetes, the algorithm was extensively evaluated in edge environments, demonstrating superior SLA compliance and resource efficiency compared to existing solutions.

\subsection{Experimental Setup}

The experimental evaluation of the hybrid autoscaler was conducted using a micro-service architecture deployed on Kubernetes, leveraging its configurability and suitability for edge deployment scenarios. The DeathStarBench social network application was chosen as the testbed, simulating large-scale social network functionalities such as user registration, timeline management, and content interaction. The architecture incorporated NGINX as the load balancer and MongoDB for persistent storage. The design was structured with distinct logic and storage layers, allowing for the realistic simulation of dynamic workloads typical in real-world edge deployments.

\textbf{Workload Type and Generation:} The workload represented IoT application usage patterns with daily cycles, including peaks in the morning and evening. These patterns were emulated using the \textit{wrk2} HTTP generator, which was customized with an algorithm introducing randomness to reflect real-world variations. The requests focused on the \textit{home-timeline-service}, a critical component responsible for retrieving user timelines. Approximately 2.55 million GET requests were generated over five days, ensuring adequate data for a comprehensive evaluation of the autoscaler's performance.

\textbf{Cluster Configuration:} The cluster consisted of six virtual machines deployed on the Melbourne Research Cloud, configured into cloud and edge layers. The cloud layer included two high-resource VMs for the Kubernetes control plane and database storage, while the edge layer comprised four low-resource VMs to simulate the resource constraints typical of edge deployments. Simulated latency was added between the cloud and edge layers to represent geographical separation. Kubernetes v1.28.2 was used with Flannel as the inter-pod communication layer, and Prometheus, integrated with Jaeger, provided system monitoring and latency tracking.

\textbf{Baseline Algorithms:} Two baseline autoscalers were compared to the hybrid solution:
\begin{itemize}
\item \textbf{Default Kubernetes \textcolor{green}{Horizontal Pod Autoscaler (HPA)}:} Reactive scaling with no workload awareness, configured with a 50\% CPU utilization threshold. It applied round-robin scheduling and suffered from significant cold start delays.
\item \textbf{\textcolor{green}{Hybrid Autoscaler}:} The proposed solution combines reactive and proactive strategies, employing an LSTM-based forecaster trained on noise-reduced, pre-processed time-series data for anticipatory scaling, complemented by reactive scaling for immediate response.
\end{itemize}

\subsection{Results}

The \textit{home-timeline-service} workload evaluation demonstrated the hybrid autoscaler’s superiority in maintaining service quality under dynamic conditions, outperforming the default algorithm in request latency and SLA compliance.

\begin{figure}
\centering
\begin{subfigure}{0.8\textwidth}
\centering
\includegraphics[width=\textwidth]{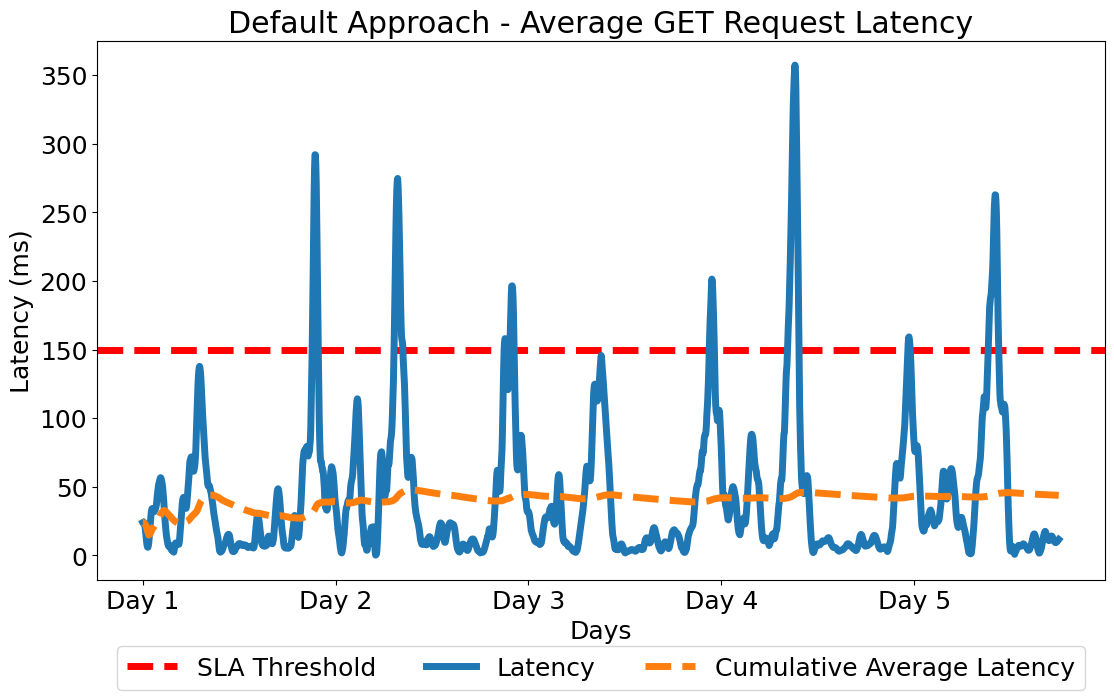}
\caption{Request latency using Default HPA.}
\label{fig:default_latency}
\end{subfigure}
\vspace{0.3cm} 

\begin{subfigure}{0.8\textwidth}
\centering
\includegraphics[width=\textwidth]{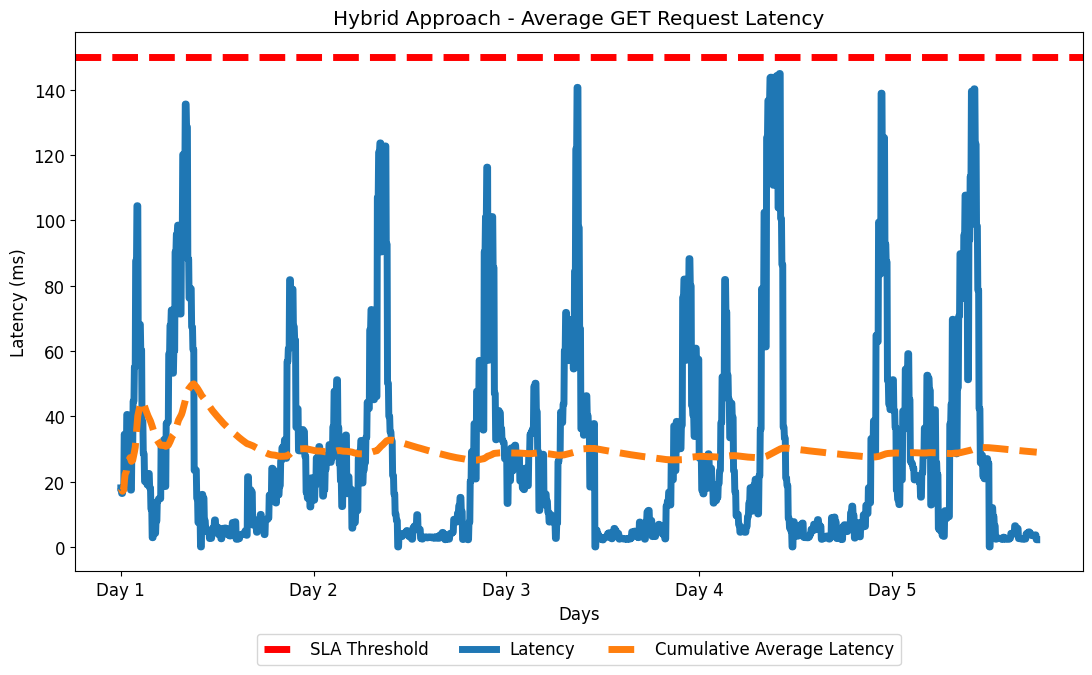}
\caption{Request latency using Hybrid Autoscaler.}
\label{fig:hybrid_latency}
\end{subfigure}

\caption{Comparison of request latency between the Default HPA and Hybrid Autoscaler under dynamic workloads.}
\label{fig:latency_comparison}
\end{figure}

\textbf{Request Latency:} Latency analysis revealed the hybrid autoscaler’s capability to eliminate cold start delays, a critical limitation of reactive strategies. The average latency remained under 30 ms, well within the flexible SLA threshold of 150 ms. In contrast, the default HPA exhibited frequent latency spikes exceeding 300 ms. Figure \ref{fig:latency_comparison} compares the request latencies of the hybrid and default autoscalers, highlighting the hybrid solution’s consistent performance.

\section{Current Limitations}
\subsection{Resource-Heavy Deployment}
Several auto-scaling solutions discussed here have been created for cloud computing architectures. Thus deploying them to an edge architecture poses further challenges, as they may occupy large portions of the limited available resources. When this occurs, it becomes more difficult to maintain SLA compliance while keeping deployment costs down.
\subsection{Custom Metrics Support}
A lot of the algorithms are hard-coded to use certain metrics to perform auto-scaling, for example, CPU and memory usage. While this makes the algorithm easier to deploy, it restricts its customization capabilities, since different deployments may wish to autoscale on different user metrics such as I/O load.
\subsection{Lack of Initial Training Data}
Proactive autoscalers rely on large amounts of training data to accurately predict the resource requirements for the deployment. This poses an issue when the autoscaler is initially deployed, as such data may not exist. During this period, the autoscaler is not guaranteed to output accurate results, damaging its SLA compliance and thus making it unsuitable for edge computing. Hybrid autoscalers attempt to remedy this issue by falling back to the reactive component while the initial data is being generated.
\subsection{Complex Hyper-parameter Tuning}
\begin{sloppypar}
A problem exclusive to proactive and hybrid autoscalers are the hyper-parameters which come bundled with the algorithms. These hyper-parameters require delicate tuning before they can be deployed in a real world setup. Failure to do so may result in erroneous results, jeopardizing the SLA compliance of the deployment.
\end{sloppypar}
\section{Future Work}
\subsection{Multi-variate Auto-scaling}
The forecasters used in most proactive and hybrid autoscalers are a uni-variate model. This means that at any given time, only one variable is changing. Therefore the input is a one-dimensional array. An extension can be made to the forecaster to convert it into a multi-variate model. In this approach, multiple variables are being modified at a given time, thus the input to the model is a more complex two-dimensional array.\par

The benefit of a multi-variate approach is that a more holistic auto-scaling decision can be made using this. The drawback of such an approach is that the proactive forecaster training times will significantly increase due to the additional complexity of the input as well as output.

\subsection{Multi-SLA Constraints}
Another enhancement which can be made is to support the use of multiple SLA constraints. By modifying the autoscaler into a multi-parameter model, it can keep a track of several metrics, and if a violation occurs on any one of them, can tune the auto-scaling process accordingly.\par

Such a modification opens up several other possibilities too. Each SLA metric can be given its own weight to denote importance. The current hybrid autoscaler only considers a single action of tuning the forecaster hyper-parameter variables on SLA violation occurrences. By making it a multi-SLA model with different weights attached, the autoscaler can intelligently decide what actions to take when on different scenarios such as urgent and mild SLA violations.\par

This adds further complexity to the autoscaler however, and the constant tuning of forecast parameters may cause issues such as a drop in prediction accuracy and under-fitting / over-fitting. Thus a balance needs to be struck in the actions to be taken on SLA constraints being violated, and additional cooldowns need to be implemented on the frequency of such actions being implemented.

\section{Summary and Conclusions}
\label{ch:conclusion}


Autoscaling on edge computing architectures remains a major research area, due to the resource and time constraints involved. In this chapter, we describe edge computing architecture, how it differs from general cloud computing paradigms, and how these architectures may be implemented using Kubernetes. Furthermore, we describe service level agreements and how they are implemented in edge computing. We have also shown that several reactive, proactive, and hybrid autoscalers exist for cloud computing environments. However, these algorithms do not always work well in edge computing scenarios, due to a variety of reasons. Each of these algorithms and their corresponding limitations are presented in this chapter. Finally, we review the future areas of research such as multi-variate solutions which may improve the performance of such algorithms and encourage their adoption industry-wide.
\par

\subsubsection*{Acknowledgments.} We thank the editors and reviewers for their insightful comments and suggestions on improving this chapter.

\section*{Author Biographies}

\parbox[t]{\linewidth}{
\noindent\parpic{\includegraphics[height=1.5in,width=1in,clip,keepaspectratio]{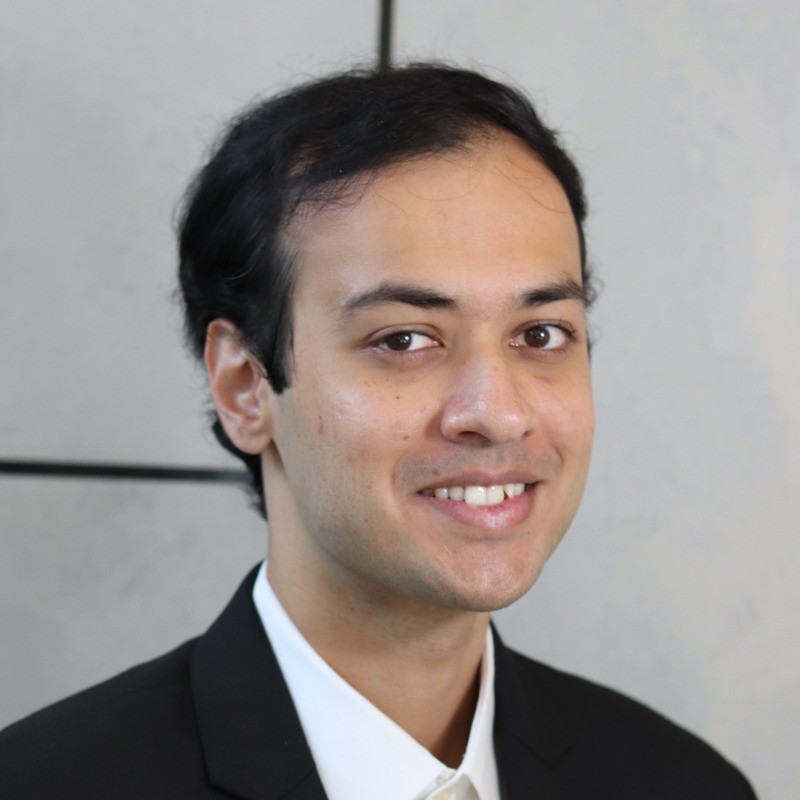}}
\noindent {\bf Suhrid Gupta} currently works as a research assistant at the University of Melbourne. He received his Master of Computer Science degree from the University of Melbourne, with his primary research interests in the field of edge computing, auto-scaling algorithms, and machine learning.
}
\vspace{4\baselineskip}

\par\noindent
\parbox[t]{\linewidth}{
\noindent\parpic{\includegraphics[height=1.5in,width=1in,clip,keepaspectratio]{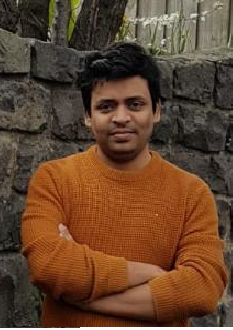}}
\noindent {\bf Muhammed Tawfiqul Islam} is a Lecturer in the School of Computing and Information Systems (CIS) at the University of Melbourne, Australia. He completed his PhD from the School of Computing and Information Systems (CIS) at the University of Melbourne in 2021. His research interests include resource management, cloud computing, big data, stream computing, and software-defined networks.
}
\vspace{4\baselineskip}

\par\noindent
\parbox[t]{\linewidth}{
\noindent\parpic{\includegraphics[height=1.5in,width=1in,clip,keepaspectratio]{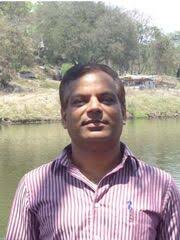}}
\noindent {\bf Rajkumar Buyya} is a Redmond Barry Distinguished Professor and Director of the Cloud Computing and Distributed Systems (CLOUDS) Laboratory at the University of Melbourne, Australia. He is also serving as the founding CEO of Manjrasoft, a spin-off company of the University commercializing its innovations in Cloud Computing. He has authored over 750 publications and four textbooks. He is one of the highly cited authors in computer science and software engineering worldwide (h-index 169 with 152,400+ citations). He is among the world’s top 2 most influential scientists in distributed computing in terms of both single-year impact and career-long impact based on a composite indicator of the Scopus citation database. He served as the founding Editor-in-Chief (EiC) of IEEE Transactions on Cloud Computing and is now serving as EiC of the Journal of Software: Practice and Experience.
}
\vspace{4\baselineskip}

\end{document}